\documentclass[%
 reprint,
 amsmath,amssymb,
 aps,
prb,
]{revtex4-2}

\usepackage{subcaption}
\usepackage{graphicx}
\usepackage{dcolumn}
\usepackage{xcolor}
\usepackage{comment}
\usepackage[colorlinks=true,linkcolor=blue,urlcolor=blue,citecolor=blue]{hyperref}
\usepackage{float}
\usepackage{bm}

\begin{document}

\title{Cluster Ising quantum batteries can mimic super-extensive charging power}

\author{Anna Pavone}
\thanks{These authors contributed equally to this work.}
\affiliation{Dipartimento di Fisica, Università degli studi di Genova, Via Dodecaneso 33, Genova, 16146, Italia}
\author{Federico Luigi Cavagnaro}
\thanks{These authors contributed equally to this work.}
\affiliation{Dipartimento di Fisica, Università degli studi di Genova, Via Dodecaneso 33, Genova, 16146, Italia}
\author{Matteo Carrega}
\affiliation{CNR-SPIN, Via Dodecaneso 33, 16146 Genova (Italy)}

\author{Riccardo Grazi}
\affiliation{Dipartimento di Fisica, Università degli studi di Genova, Via Dodecaneso 33, Genova, 16146, Italia}
\affiliation{CNR-SPIN, Via Dodecaneso 33, 16146 Genova (Italy)}

\author{Dario Ferraro}
\affiliation{Dipartimento di Fisica, Università degli studi di Genova, Via Dodecaneso 33, Genova, 16146, Italia}
\affiliation{CNR-SPIN, Via Dodecaneso 33, 16146 Genova (Italy)}

\author{Niccolò Traverso Ziani}
\email{niccolo.traverso.ziani@unige.it}
\affiliation{Dipartimento di Fisica, Università degli studi di Genova, Via Dodecaneso 33, Genova, 16146, Italia}
\affiliation{CNR-SPIN, Via Dodecaneso 33, 16146 Genova (Italy)}

\begin{abstract}
Quantum batteries—miniaturized devices able to store and release energy on demand—are promising both because their intrinsic energy and time scales can match those of other quantum technologies and due to the intriguing possibility of achieving super-extensive charging power. While this enhanced scaling is known to appear in several settings, it is generally believed to be forbidden in Jordan–Wigner integrable spin chains charged via quantum-quench protocols. Here, we show that an extended cluster-Ising model, despite belonging to the above category, exhibits super-extensive charging power over wide ranges of system sizes, reaching up to a thousand spins, in proper parameter regimes. This remarkable anomalous scaling is due to a corresponding super-extensive growth of the stored energy, implying that it \textcolor{black}{is limited to a large but finite size of the system} and cannot persist in the thermodynamic limit. This phenomenon appears robust against finite-temperature effects.
\end{abstract}

\maketitle
\section{Introduction}

The quest to determine whether quantum technologies can outperform their classical counterparts is one of the main challenges across modern quantum science~\cite{Ezratty2024, AguadoLNP2024}. Within this vast landscape, quantum batteries (QBs), devices devoted to store and release energy exploiting purely quantum effects, offer a fertile ground 
to explore this question, as they represent a paradigmatic setting in which energetic resources, charging mechanisms and quantum 
correlations interplay in a fruitful and nontrivial way~\cite{QuachJoule2023, CampaioliRMP2024, CamposeoAM2025, FerraroNRP2026}. 

Among the various figures of merit considered to characterize the performance of these devices, a crucial role is played by the stored 
energy and the average charging power, namely the energy which can be delivered to the battery per unit time~\cite{AndolinaPRB2018}. In multipartite architectures, \textcolor{black}{where the QB is composed} of $N$ interacting subsystems, the stored energy is typically an extensive quantity \textcolor{black}{which scales linearly with $N$}. Yet, significant interest sparked from the possibility that the charging power per subsystems might scale in a super-extensive manner, namely $\propto N^{\alpha}$ with $\alpha>0$. This behavior indicates that adding cells \textcolor{black}{can} accelerate the charging process in a way not achievable by independent units.~\cite{BinderNJP2015, CampaioliPRL2017, GaoPRR2022, HuPRL2026, Sreeram26, Schmid26}. \textcolor{black}{The possibility to achieve a considerable charging power in systems of relative small size may have a relevant impact in technological applications such as photovoltaic devices~\cite{HymasArXiv2025}}. 

Such enhanced scaling may originate from mechanisms that can be collective in nature, which can emerge also in classical or semiclassical systems, or, more interestingly, be due to \textcolor{black}{genuine quantum correlations}~\cite{AndolinaPRB2019, JuliaFarrePRR2020, FrancicaPRE2022, GyhmAVS2024, HokkyoPRL2025, CavaliereArXiv2025, CavaliereComm2025}. Notable examples within the first category are the so called Dicke QBs~\cite{FerraroPRL2018, GemmeBatteries2023, QuachSciAdv2022, CarrascoPRE2022, WangPRA2024, YangPRB2024, CanzioPRA2025, TibbenPRXEnergy2025, KurmanPRX2026, HymasArXiv2025}, where independent two-, or more generally few-, level systems are charged via the interaction with photons trapped into a resonant cavity. \textcolor{black}{Here, a super-extensive scaling of the charging power has been theoretically  predicted~\cite{FerraroPRL2018} and experimentally observed~\cite{QuachSciAdv2022, HymasArXiv2025}. This is, strictly speaking, a finite-size effect that disappears when the thermodynamic limit is properly taken into account, namely when the volume of the cavity is enhanced by increasing $N$~\cite{JuliaFarrePRR2020}.}

The emergence of a genuine quantum advantage has been theoretically predicted for the Sachdev–Ye–Kitaev QBs, where a random non-local interaction among spins can be exploited to reach a super-extensive power~\cite{RossiniPRL2020, RosaJHEP2020} and for properly engineered harmonic oscillator QBs with non-linear charger-battery coupling~\cite{AndolinaPRL2025}.

In spin based QBs~\cite{LePRA2018,RossiniPRB2019,GhoshPRA2020, ArjmandiPRE2022, CatalanoPRXQ2024, Chand26} \textcolor{black}{the super-extensive scaling in the power is bounded by} both the range of interaction and the coordination within the lattice formed by the spins~\cite{CampaioliPRL2017, JuliaFarrePRR2020, GyhmPRL2022, AliArXiv2024}. In this context, {\color{black} Jordan-Wigner} integrable spin chain QBs have had a major impact in determining robust features of the charging dynamics in the thermodynamic limit~\cite{GraziPRL2024, GraziCSF2025}. However, since they can be mapped into free fermions via \textcolor{black}{Jordan-Wigner} transformation~\cite{FranchiniBook2017}, \textcolor{black}{they are generally believed not to show super-extensive scaling}~\cite{JuliaFarrePRR2020} {\color{black} even in cases where general bounds would allow for it\cite{GraziCSF2025}.}

In the present paper we will demonstrate that, \textcolor{black}{while this is indeed the case in the thermodynamic limit, a finite-size super-extensive scaling can emerge in an extended cluster-Ising model~\cite{SmacchiaPRA2011, DingPRE2019, KheiriSR2025, Puri25}. By considering different scaling regimes of the cluster interaction, we identify regions of parameter space where the maximal power exhibits apparent super-extensive scaling with the system size. We further show that this behavior is already encoded in the short-time dynamics, where it can be traced back to the scaling of the initial growth rate of the charging power. To gain physical insight into this phenomenon, we analyze the structure of fermionic occupation numbers, which reveal how the range of interactions and the number of sites play a role in the effect. Finally, we show that such behavior does not occur in general: an alternative integrable cluster–Ising model lacking the specific multi-string structure does not display finite-size super-extensive scaling.}

\textcolor{black}{The paper is organized as follows. In Section \ref{sec:model} we introduce the generalized cluster–Ising model and its exact solution via Jordan–Wigner transformation and Fourier decomposition. Section \ref{sec:battery} presents the charging protocol and the main figure of merit, namely the average charging power. The numerical analysis of the scaling behavior with respect to the system size and interaction range is reported in Section \ref{sec:results}, where we identify parameter regimes exhibiting finite-size super-extensive growth. In Section \ref{sec:discussion}, we complement these results with a semi-analytical study of the short-time regime, showing that signatures of the same finite-size enhancement are already visible in the initial slope of the charging power. We further analyze the role of fermionic occupation numbers, which provide insight into the mechanism underlying the observed scaling behavior. Finally, in Section \ref{sec:unfavorable} we discuss an alternative integrable cluster–Ising variant that does not display super-extensive scaling. The conclusions are drawn in Section \ref{conclusions}.
}

{\color{black}\section{\label{sec:model}Model}
As QB Hamiltonian we consider a generalized cluster-Ising model~\cite{SmacchiaPRA2011, DingPRE2019, KheiriSR2025}.}
Explicitly we set (up to an overall energy prefactor)
\begin{widetext}
\begin{eqnarray}
H&=&- \cos\phi \sum_{j=1}^N \sum_{l=1}^n \frac{1}{n}
\sigma^{x}_{j}\, O^{z}_{j,l}\, \sigma^{x}_{j+l+1}
+ \sin\phi \sum_{j=1}^N 
\sigma^{z}_{j}, \label{eq:H2}
\end{eqnarray}
\end{widetext}
with
\begin{equation} \label{Ozjn}
O^{z}_{j,n} = \prod_{k=1}^{n} \sigma^{z}_{j+k}.
\end{equation}
Here, $\sigma^{\alpha}_j$, ($\alpha=x,y,z$) are the Pauli matrices in the usual representation for the site $j$, {\color{black} $\phi$ is a free real parameter}, $N$ is the number of sites composing the chain and $n<N$ is the range of the cluster interaction term. We adopt periodic boundary conditions. {\color{black} The Hamiltonian under investigation is characterized by multiple spin} strings of lengths 3,4,..,$n+1$, all with the same coefficient in front. \textcolor{black}{In analogy with what usually done in the Dicke QB case~\cite{JuliaFarrePRR2020}}, the $1/n$ term is introduced to have extensive energies in the thermodynamic limit.\\
{\color{black}The model is} \textcolor{black}{Jordan-Wigner} integrable, meaning that {\color{black}it} can be mapped into a free fermion system~\cite{FranchiniBook2017}. It is worth noting that a model with different weights in front of each string would also be integrable, but {\color{black}  we do not consider this case in the following}. Explicitly, to solve the model, one first performs the \textcolor{black}{Jordan-Wigner} transformation to spinless fermions $c_j$ given by~\cite{JordanWigner1928}
\begin{eqnarray}
\sigma^z_j&=&1-2 c^\dag_j c_j,\\
\sigma^{+}&=&\prod_{l<j}(1-2 c^\dag_l c_l) c_j,\\
\sigma^{-}&=&\prod_{l<j}(1-2 c^\dag_l c_l) c^\dag_j,
\end{eqnarray}
with $\sigma^{\pm}_j=(\sigma^x_j\pm i \sigma^y_j)/2$. Subsequently, to exploit the translational invariance, it is useful to \textcolor{black}{consider} the Fourier expansion ($q=0,1,..,N-1$)
\begin{equation}
\psi_q =\frac{1}{\sqrt{N}} e^{-i\frac{\pi}{4}} \sum_{j=1}^{N} e^{-i \frac{2\pi}{N} q j} c_{j}
\end{equation}
to get the Bogoliubov-de Gennes-like \textcolor{black}{Hamiltonian}
\begin{eqnarray} \label{Ham_Sum_To_Diag}
H&=&\sum_q(\psi^\dag_q,\psi_{-q})\mathcal{H}^{q}(\psi_q,\psi^\dag_{-q})^T
\end{eqnarray}
with
\begin{equation} \label{Ham_With_Tau_Matrices}
\mathcal{H}^{q}=A^q\tau^z+C^q\tau^x,
\end{equation}
where $\tau^{z/x}$ are Pauli matrices in a fictitious two-dimensional space and
\begin{align}
A^q
&= \frac{\cos\phi}{n}\!\left[
\frac{\sin\!\big(\tfrac{(n+1)\theta_q}{2}\big)\,\cos\!\big(\tfrac{(n+2)\theta_q}{2}\big)}{\sin(\theta_q/2)}
-\cos\theta_q
\right]
-\sin\phi \notag \\
C^q
&= -\,\frac{\cos\phi}{n}\!\left[
\frac{\sin\!\big(\tfrac{(n+1)\theta_q}{2}\big)\,\sin\!\big(\tfrac{(n+2)\theta_q}{2}\big)}{\sin(\theta_q/2)}
-\sin\theta_q
\right]
\label{eq:Aq2Cq2}
\end{align}
with $\theta_q = \frac{2\pi}{N}q.$ {\color{black} The calculation leading to Eq. (\ref{eq:Aq2Cq2}) is lengthy but basically identical to the one characterizing the XY chain, reported with great detail for example in \cite{FranchiniBook2017}. The only qualitatively different steps are realizing that the $O^{z}_{j,n}$ either cancel with the \textcolor{black}{Jordan-Wigner} string or contribute to the factorization of the fermion parity in the boundary terms, depending on whether the spin operators $\sigma_j^{x}$ and $\sigma_{j+l+1}^{x}$ belong to the same range $\left[1,N\right]$ or not. Moreover, in our derivation, we have restricted the analysis to the odd fermion parity sector, a choice that does not qualitatively alter the results for two reasons. i) The Hamiltonian commutes with the fermion parity operator and hence the dynamics, which we will consider in the following, is restricted to the same fermion parity sector as the initial state. ii) We have checked that the same physics occurs also in the even parity sector. The independence of our results on the fermion parity sector is expected \textit{a priori} since the observables we are interested in are extensive, and hence even potential boundary effects, related for instance to topological frustration, are not expected to be detectable~\cite{tf1,tf2,tf3,tf4}}.\\

\section{\label{sec:battery}{Battery setup}}
The charging of this QB is implemented via a time-dependent protocol. Specifically, the Hamiltonian becomes time-dependent, $H(t)$, through stepwise changes in the angle $\phi$, effectively realizing a double quantum quench~\cite{MitraARCMP2018, PortaSR2020}. Explicitly
{\color{black}
\begin{equation}
\phi(t)=\phi_B(\Theta(-t)+\Theta(t-\tau))+\phi_C \Theta(t)\Theta(-t+\tau),
\end{equation}
where $\phi_{B/C}$ denote respectively the battery and charging parameters}, $\tau$ is the charging time and $\Theta(\cdot)$ is the Heaviside step function. 
Correspondingly, one has
{\color{black}
\begin{eqnarray}
A^q(t)&=&A_B^{q}(\Theta(-t)+\Theta(t-\tau))+A_{C}^{q}\Theta(t)\Theta(-t+\tau),\nonumber\\
C^q(t)&=&C_{B}^{q}(\Theta(-t)+\Theta(t-\tau))+C_{C}^{q}\Theta(t)\Theta(-t+\tau)\nonumber.
\end{eqnarray}}
For $t<0$ the system is prepared either in the ground state or in a thermal state with temperature $T=1/(k_B\beta)$, with $k_B$ denoting the Boltzmann constant.

The fundamental figures of merit to characterize this QB is the energy transferred per spin during the time evolution, referred to as the charging energy. Explicitly, we have
\begin{equation}
E(\tau) = \frac{
{Tr}\!\left[
e^{-\beta H(0^{-})}
\left(
H_{H}(\tau^{+}) - H_{H}(0^{-})
\right)
\right]
}{N\,\,
{Tr}\!\left[e^{-\beta H(0^{-})}\right]
},
\label{eq:DeltaE}
\end{equation}
where $\operatorname{Tr}[\cdot]$ denotes the trace and $H_{H}(t)$ is the Hamiltonian in the Heisenberg representation. 
The calculation of $E(\tau)$ can be performed explicitly {\color{black}(see Appendix \ref{App:Energy_Chaos} for more details)} and reads as
\begin{widetext}
\begin{equation}
E(\tau)=\sum_{q}
\frac{1 - \cos\!\big( 2\,\omega(q)\,\tau \big)}{2\,N\,\varepsilon(q)\,\omega^{2}(q)}
\,
\left(C_B^{q}\, A_C^{q}- A_B^{q}\,C_C^{q} \right)^{2}
\,
\tanh\!\left( \frac{\beta\,\varepsilon(q)}{2} \right). \label{Energy_Chaos}
\end{equation} 
\end{widetext}

Here,
\begin{eqnarray}
\varepsilon(q)&=&\sqrt{\left(A_B^{q}\right)^2+\left(C_B^{q}\right)^2},\\
\omega(q)&=&\sqrt{\left(A_C^{q}\right)^2+\left(C_C^{q}\right)^2}
\end{eqnarray}
are the energy dispersions before and during the charging respectively. Notice that the energy \textcolor{black}{per spin does not scale with $N$, indicating an extensive behaviour} in the thermodynamic limit.\\

The average charging power $P(\tau)$ follows directly from the energy and is given by
\begin{equation}
P(\tau)=E(\tau)/\tau.
\end{equation}
For the following analysis, it is useful to introduce the maximal figures of merit. The maximal charging power $P_M$ is defined as
\begin{equation}
P_M= max_\tau P(\tau)
\end{equation}
where the value of $\tau$ corresponding to $P_M$ is indicated as $\tau_{M}$. 
\\

{\color{black}\section{\label{sec:results}{Results}}

We show that, by appropriately choosing the range of $N$ and the functional dependence $n(N)$, it is possible to obtain super-extensive scaling of the maximal charging power $P_M$ within a finite range of system sizes. Moreover, this finite-size super-extensive behavior can persist over broad intervals of $N$, does not require fine tuning, and is robust against variations in temperature. It is important to emphasize that this finite-size super-extensive scaling does not persist in the thermodynamic limit $N \rightarrow \infty$, in agreement with Eq.~\ref{Energy_Chaos}.

Fig.~1(a) shows the charging power $P(\tau)$ as a function of the charging time $\tau$ for different values of $N$, and $nN \simeq 2000$. The remaining parameters, whose specific values do not qualitatively affect the physics, are set to $T = 0$, $\phi_B = 0.6$, and $\phi_C = \pi/4$. The key feature to note is that the maximal charging power $P_M$ increases with increasing $N$, signaling finite-size super-extensive scaling over the range considered. In addition, $P_M$ is always reached at approximately the same value of $\tau$, indicating that the observed scaling originates from an increase in the stored energy rather than from a speedup of the charging process.

Fig.~1(b) reports a fit of the maximal charging power extracted from panel (a) using the scaling form $P_M = a N^\alpha$. Remarkably, the fitted exponent $\alpha = 0.86$ is clearly distinct from zero, confirming the super-extensive character of the scaling. To better address this point, we now clarify that, for each favorable scaling regime, we have explored system sizes beyond those shown in the figures. We have verified that, upon further increasing N, the apparent super‑extensive behavior crosses over and eventually disappears, consistently with thermodynamic expectations. The ranges of N displayed in the manuscript are chosen to highlight this finite‑size super‑extensive regime.

\begin{figure}[H]
    \centering
    \begin{subfigure}{0.48\textwidth}
        \centering
        \includegraphics[width=0.68\linewidth]{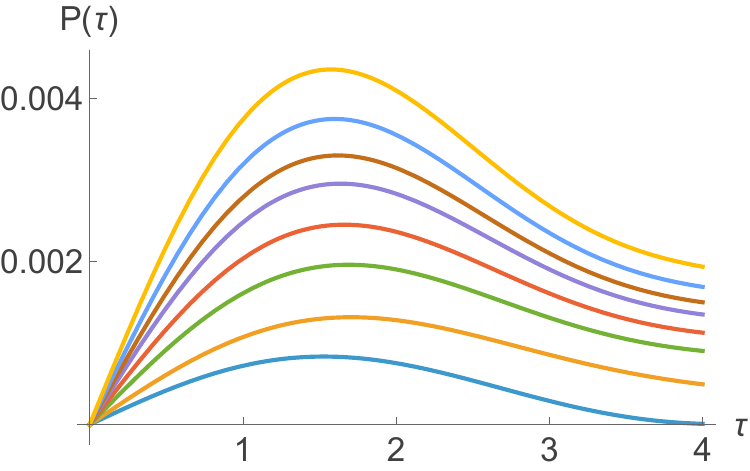}
        \caption{}
    \end{subfigure}
    \hfill
    \begin{subfigure}{0.48\textwidth}
        \centering
        \includegraphics[width=0.72\linewidth]{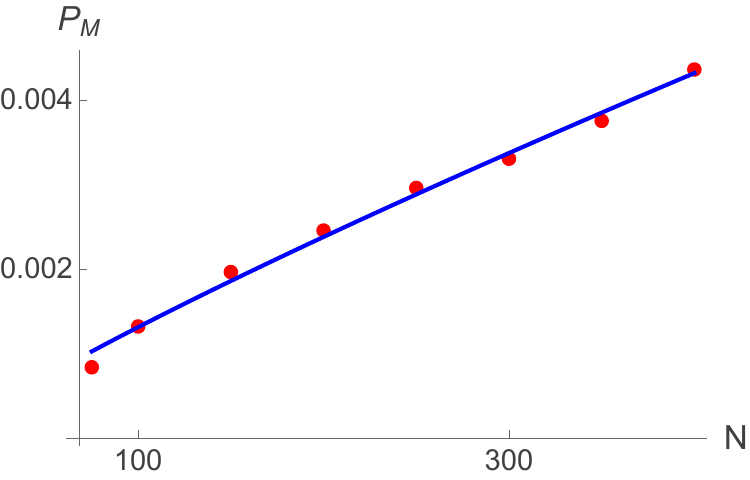}
        \caption{}
    \end{subfigure}
    \caption{(a) Charging power $P(\tau)$ as a function of $\tau$ for values of $N$ ranging from 75 (dark blue curve) to 400 (yellow curve), with $\phi_B = 0.6$, $\phi_C = \pi/4$, and scaling $nN = 2000$. (b) Maximal charging power $P_M$ as a function of $N$ for the case shown in the above panel (red dots), together with the fit curve $P_M = a N^\alpha$ (blue line). The best-fit parameters are $a = 2.5 \cdot 10^{-5}$ and $\alpha = 0.86$.}
    
    \label{Fit_nN}
\end{figure}

\begin{figure*}[t]
    \centering
    \begin{subfigure}{0.48\textwidth}
        \centering
        \includegraphics[width=0.8\linewidth]{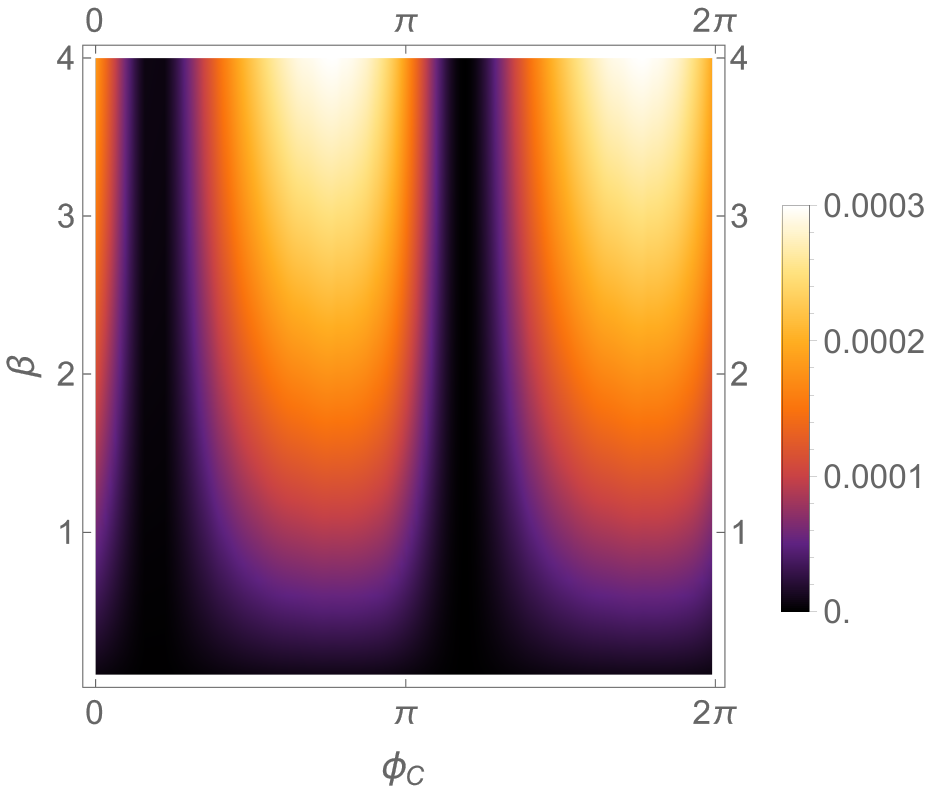}
        \caption{}
    \end{subfigure}
    \hfill
    \begin{subfigure}{0.48\textwidth}
        \centering
        \includegraphics[width=0.8\linewidth]{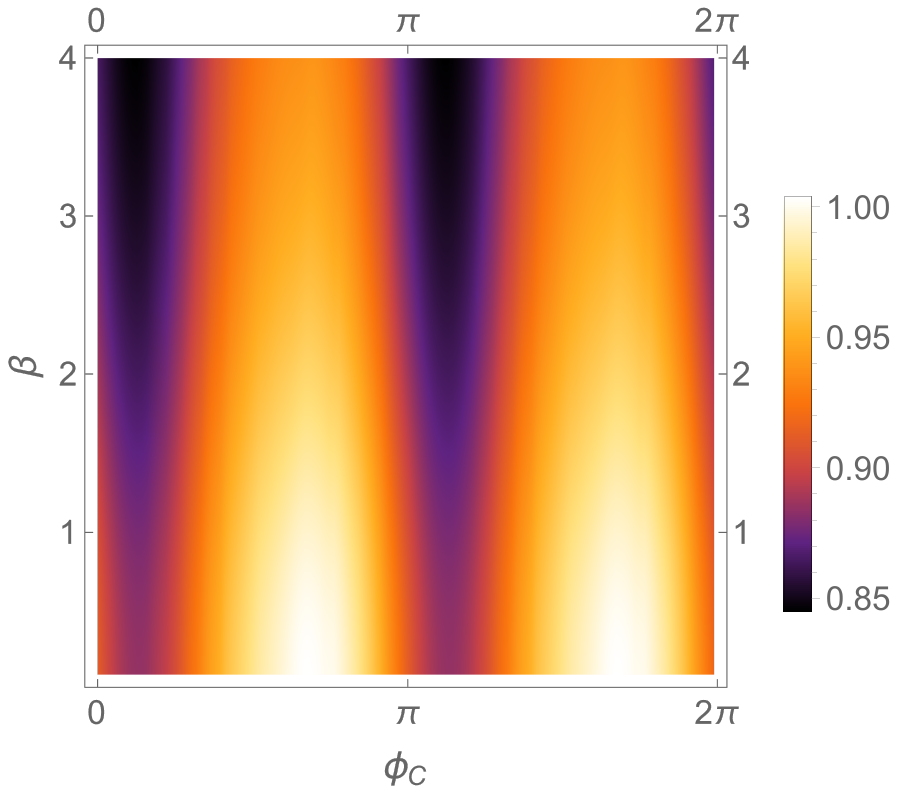}
        \caption{}
    \end{subfigure}
    
    \caption{Density plots of the fit coefficients $a$, panel (a), and $\alpha$, panel (b), in the scaling law $P_M =a N^{\alpha}$, used to describe the maximum power as a function of $N$, showing their dependence on $\phi_C$ and the temperature $\beta$, at fixed $\phi_B = 0.6$ and $nN = 2000$.}
    
    \label{Nn_Density}
\end{figure*} 
In Fig.~2, we demonstrate the robustness of finite-size super-extensive charging with respect to variations in the angle $\phi_C$ (analogous results are obtained upon varying $\phi_B$, not shown) and the temperature. Panel (a) displays the fitted coefficient $a$. As expected, $a = 0$ for $\phi_C = \phi_B$ and $\phi_C = \phi_B + \pi$, while it remains nonzero for other values of $\phi_C$. Furthermore, $a$ decreases as the temperature increases. Panel (b) shows the behavior of the exponent $\alpha$, which, within the parameter range considered, oscillates between approximately $0.85$ and $1$.

Finally, we stress that these results are not specific to the particular choice $nN = 2000$. Similar behavior is observed for other relations between $n$ and $N$, as discussed in Appendix~B.}

{\color{black} \section{Short-time regime and occupation number interpretation} \label{sec:discussion}
In this section, we present two complementary analyses aimed at further characterizing our results and elucidating the physical mechanisms underlying the finite-size super-extensive scaling exhibited by the model.

The first analysis consists of a semi-analytical study of the short-time behavior of the charging power $P(\tau)$. The motivation for focusing on this regime stems from the results shown in Fig.~1(a). For each value of $N$, the maximal power $P_M$ depends only weakly on the precise value of the charging time $\tau$, and within the interval between $\tau = 0$ and $\tau = \tau_{M}$, the function $P(\tau)$ is approximately linear in $\tau$. As a consequence, whenever finite-size super-extensive charging power is observed, it already manifests itself in the initial slope $dP(\tau)/d\tau$ evaluated at $\tau = 0$. This criterion may prove particularly useful for detecting finite-size super-extensive behavior in systems where a full dynamical analysis is technically demanding, such as non-integrable models.

The second analysis focuses on the contributions of the different quasi-momenta $q$ to the maximal charging power $P_M$. We find that the observed super-extensive behavior most likely arises from the interplay between two factors: the number and amplitude of the oscillations of the fermionic occupation numbers—which are controlled by the parameter $n$—and the sampling of the dispersion relation, which is determined by the system size $N$. Our analysis indicates that, in order to achieve super-extensive scaling over a given range of $N$, it is advantageous to reduce the number of oscillations in the occupation numbers as $N$ increases. This insight may provide useful guidance for the construction of alternative models exhibiting finite-size super-extensive scaling.
\subsection{Short time regime} \label{subsec_short_time}
In this subsection, we develop a semi-analytical analysis of the short-time behavior of the charging power. We consider the zero-temperature limit of Eq. \eqref{Energy_Chaos} and we mainly focus on the battery and charging parameters $\phi_{B} = 0$ and $\phi_{C} = \phi$ respectively. This choice simplifies the analytical expressions without affecting the physical behavior of interest. What we show is that even at short times, the finite-size super-extensive behavior is visible. \\
By performing a first-order expansion of the cosine term in Eq. \eqref{Energy_Chaos}, we obtain
\begin{equation}
    E(\tau) \sim \sum_{q }
\frac{\tau^2}{N\,\varepsilon(q)}
\,
\left(C_B^{q}\, A_C^{q}- A_B^{q}\,C_C^{q} \right)^{2},
\end{equation}
so that
\begin{widetext}
\begin{equation}
    \frac{E(\tau)}{\tau} \sim \bigg\{\frac{1}{N}\sum_{q }
\frac{1}{\varepsilon(q)}
\,
\left(C_B^{q}\, A_C^{q}- A_B^{q}\,C_C^{q} \right)^{2}\bigg\} \tau + O(\tau^3) = \gamma \tau + O(\tau^3).
\end{equation}
\end{widetext}
From this analysis, it emerges that, to leading order, the power grows linearly in time. Our goal is to show that signatures of super-extensivity already emerge at this early stage and can be captured by analyzing the slope $\gamma$ of this linear behavior. From the definitions reported in Eq. \eqref{eq:Aq2Cq2}, we define
\begin{equation}
    \begin{aligned}
        &X_{cos} = \frac{\sin\!\big(\tfrac{(n+1)\theta_q}{2}\big)\,\cos\!\big(\tfrac{(n+2)\theta_q}{2}\big)}{\sin(\theta_q/2)}\\
        &X_{sin} = \frac{\sin\!\big(\tfrac{(n+1)\theta_q}{2}\big)\,\sin\!\big(\tfrac{(n+2)\theta_q}{2}\big)}{\sin(\theta_q/2)},
    \end{aligned}
\end{equation}
so that the coefficients $A^q_{B/C}$ and $C^q_{B/C}$ are respectively
\begin{equation}
\begin{aligned}
A_B^q
&= \frac{1}{n}\!\left[
X_{cos}
-\cos\theta_q
\right] \notag \\
C_B^q
&= -\,\frac{1}{n}\!\left[
X_{sin}
-\sin\theta_q
\right] \\
A_C^q
&= \frac{\cos\phi}{n}\!\left[X_{cos}
-\cos\theta_q
\right]
-\sin\phi \notag \\
C_C^q
&= -\,\frac{\cos\phi}{n}\!\left[X_{sin}
-\sin\theta_q
\right].
\end{aligned}
\end{equation}
From straightforward calculation we can derive
\begin{equation}
    \left(C_B^{q}\, A_C^{q}- A_B^{q}\,C_C^{q} \right)^{2} = \frac{\sin^2(\phi)}{n^2}(X_{sin}-\sin\theta_q)^2
\end{equation}
and
\begin{equation}
    \varepsilon(q) = \frac{1}{n}\sqrt{(X_{cos}-\cos\theta_q)^2 + (X_{sin}-\sin\theta_q)^2},
\end{equation}

so that
\begin{equation} \label{gamma_non_simpl}
    \gamma=\frac{\sin^2(\phi)}{nN} \sum_{q=0}^{N-1} \frac{(X_{sin}-\sin\theta_q)^2}{\sqrt{(X_{cos}-\cos\theta_q)^2 + (X_{sin}-\sin\theta_q)^2}}.
\end{equation}
We can simplify the term inside the sum by noticing that
\begin{equation}
    X_{cos} + iX_{sin} = \sum_{k=1}^{n+1} e^{ik\theta_q}
\end{equation}
so that
\begin{equation} \label{Geom_Series}
\begin{aligned}
    (X_{cos} - \cos\theta_q) + i(X_{sin} - \sin\theta_q) &= \sum_{k=2}^{n+1} e^{ik\theta_q} = \\&= e^{i\frac{(n+3)\theta_q}{2}} \frac{\sin(\frac{n\theta_q}{2})}{\sin(\frac{\theta_q}{2})}.
\end{aligned}
\end{equation}
Now, Eq. \eqref{gamma_non_simpl} can be written in terms of sines only by noticing that the numerator of the term inside the sum is the imaginary part of the complex number in Eq. \eqref{Geom_Series} squared, while the denominator is its module. The final expression for $\gamma$ can be written as
\begin{equation} \label{gamma_simpl}
    \gamma=\frac{\sin^2(\phi)}{nN} \sum_{q=0}^{N-1} \left|\frac{\sin(\frac{n\theta_q}{2})}{\sin(\frac{\theta_q}{2})}\right| \sin^2 \left(\frac{(n+3)\theta_q}{2}\right).
\end{equation}
In Fig. \ref{gamma_N_tau_corti_fav} we plot $\gamma$ as function of the number of sites $N$ for the specific case $nN = \text{cost.}$ We observe that the slope of the charging power in the short-time limit grows as $N$ ($n$) becomes bigger (smaller). In panel (a) the case discussed with $\phi_B=0$, $\phi_C=\phi$ is shown. In panel (b), to show the robustness of the effect, we show a case with $\phi_B\neq 0$. To conclude the analysis, in Fig. \ref{gamma_N_tau_corti_sfav},  we show, in panel (a) the charging power and in panel (b) $\gamma$ as a function of $N$ for the unfavorable case $n=\sqrt{N}$. Here, $\gamma$ does not grow monotonically with $N$.
\begin{figure}[h!]
    \centering
    \begin{subfigure}{0.5\textwidth}
        \centering
        \includegraphics[width=0.8\linewidth]{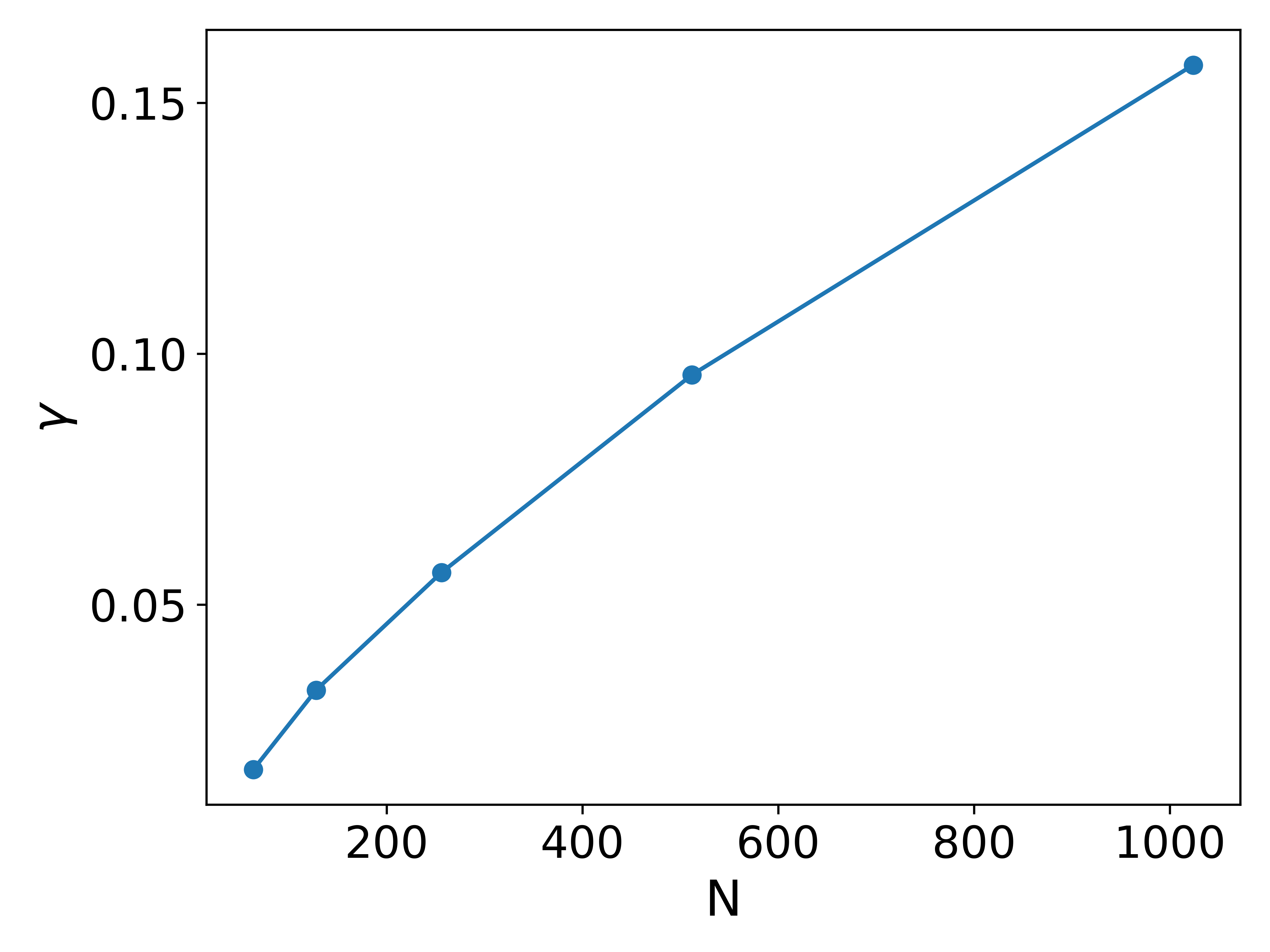}
        \caption{}
    \end{subfigure}
    \hfill
    \begin{subfigure}{0.5\textwidth}
        \centering
        \includegraphics[width=0.8\linewidth]{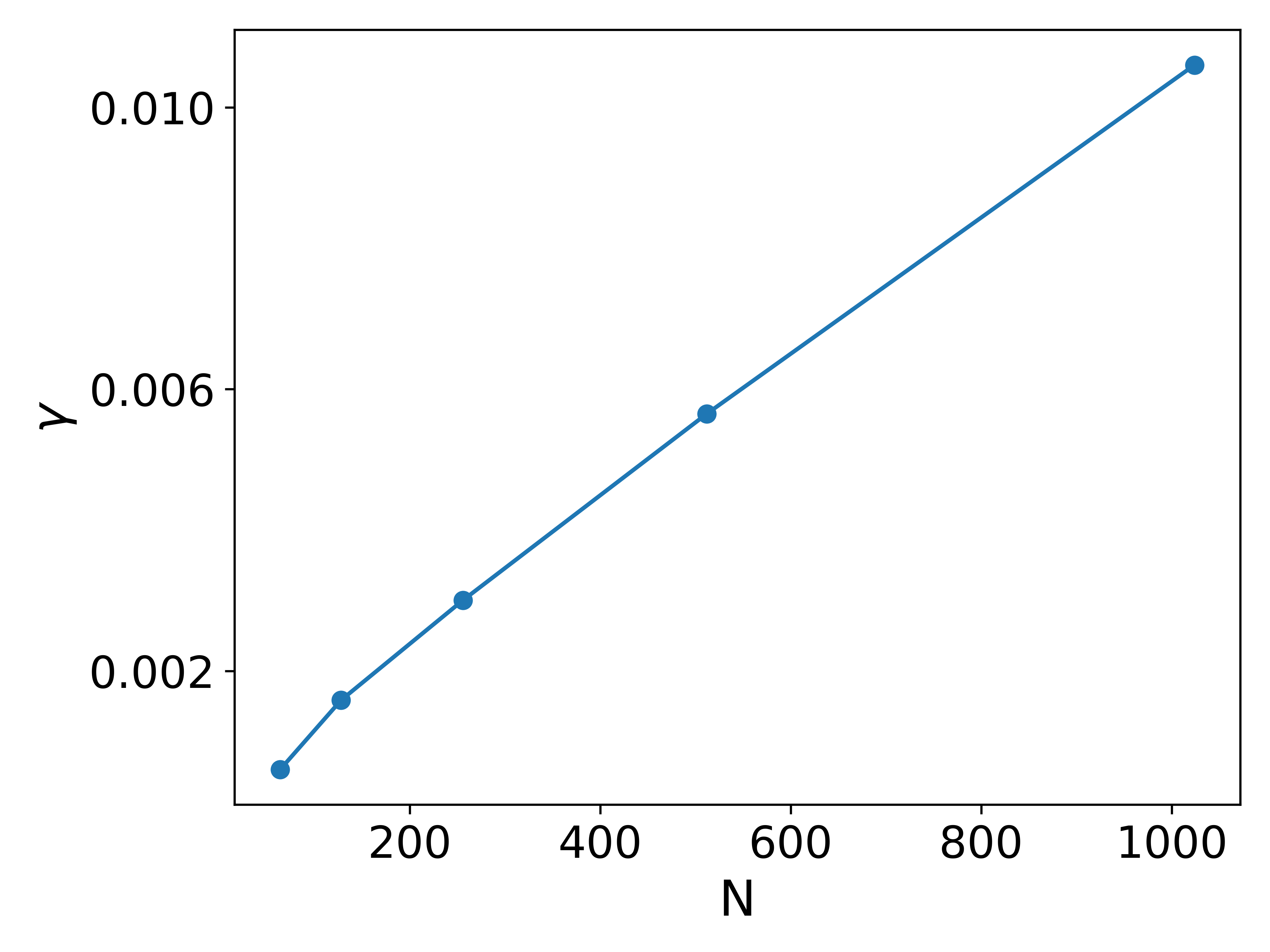}
        \caption{}
    \end{subfigure}

 \caption{Slope $\gamma$ as function of $N$ for (a) $\phi_B = 0$, $\phi_C = \pi/4$ and $nN = 2048$, (b) $\phi_B = 0.6$, $\phi_C = \pi/4$ and $nN = 2048$.}
    \label{gamma_N_tau_corti_fav}
\end{figure}

\begin{figure}[h!]
    \centering
    \begin{subfigure}{0.5\textwidth}
        \centering
        \includegraphics[width=0.75\linewidth]{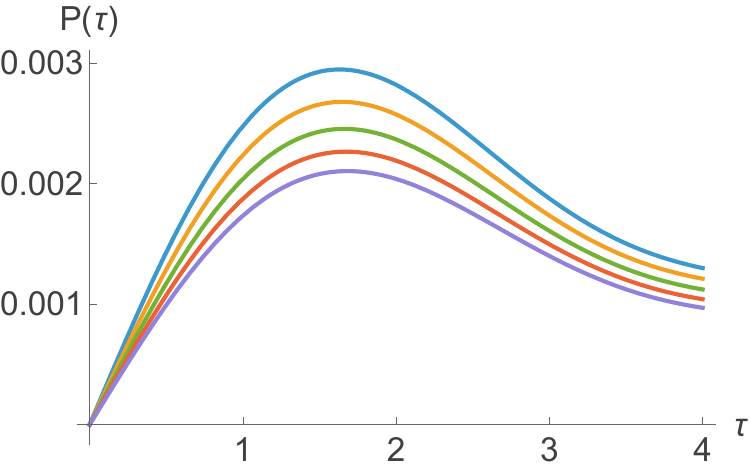}
        \caption{}
    \end{subfigure}
    \hfill
    \begin{subfigure}{0.5\textwidth}
        \centering
        \includegraphics[width=0.8\linewidth]{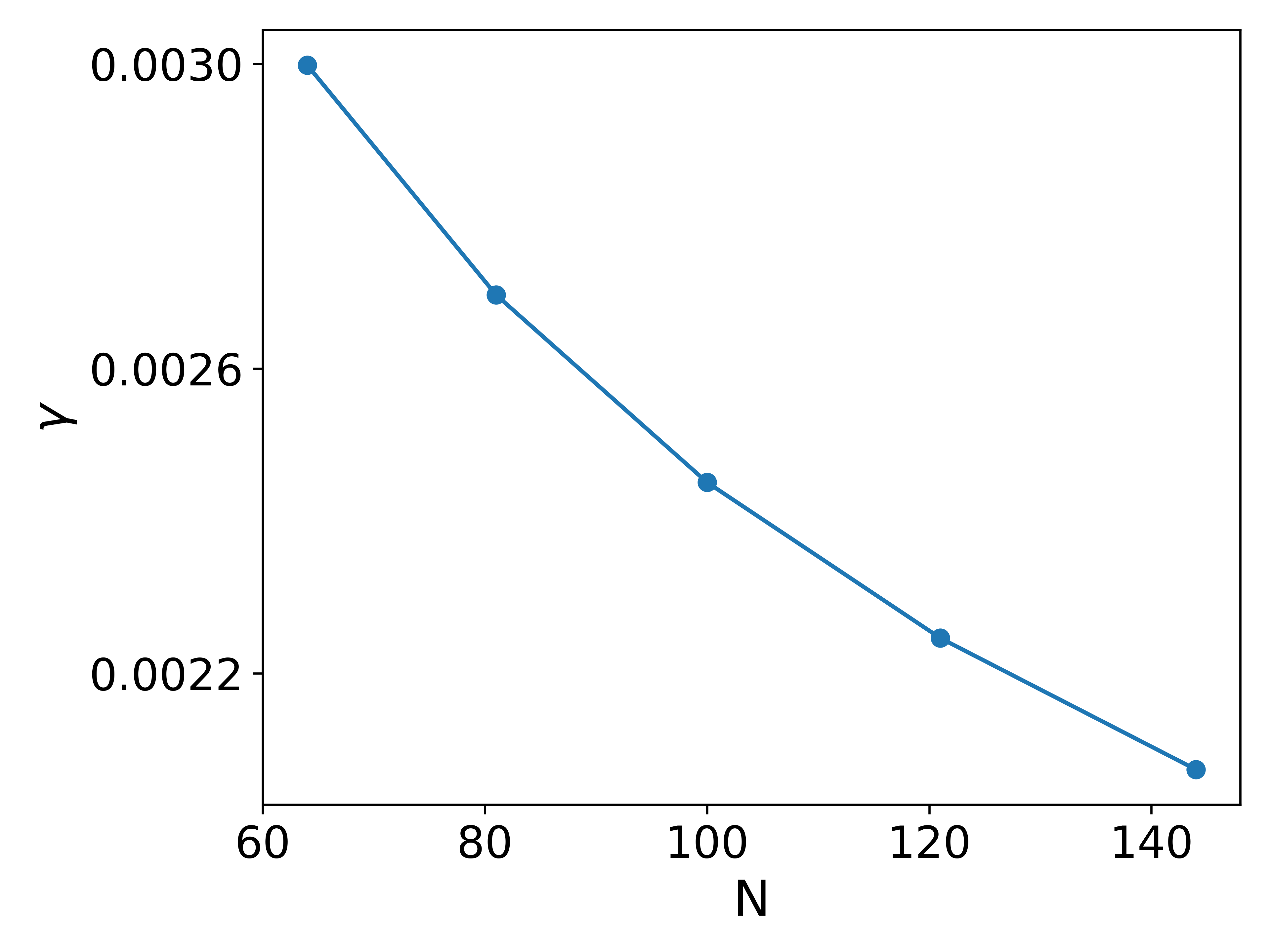}
        \caption{}
    \end{subfigure}
 \caption{(a) Charging power $P(\tau)$ as a function of $\tau$ for values of $N$ ranging from 64 (light blue curve) to 144 (purple curve), with $\phi_B = 0.6$, $\phi_C = \pi/4$, and scaling $n = \sqrt{N}$. (b) Slope $\gamma$ as function of $N$ for $\phi_B = 0.6$, $\phi_C = \pi/4$ and $n = \sqrt{N}$.}
    \label{gamma_N_tau_corti_sfav}
\end{figure}

\subsection{Study of the fermionic occupation number}

The model under consideration is translationally invariant and, at any time, its Hamiltonian can be mapped onto a set of free fermionic modes labeled by the quasi-momentum $q$. As a consequence, the stored energy appearing in Eq.~\ref{Energy_Chaos} can be written as
\begin{equation}
E(\tau)=\frac{1}{N}\sum_q \epsilon(q)\,\mathcal N_(q,\tau),
\end{equation}
where $\mathcal N_(q,\tau)$ denotes the occupation number of the fermionic mode $q$ that diagonalizes the battery Hamiltonian after a charging time $\tau$.

At zero temperature, the occupation numbers take the form
\begin{equation}
  \mathcal N(q,\tau)=  
  \frac{1 - \cos\!\big( 2\,\omega(q)\,\tau \big)}{2\,\varepsilon^2(q)\,\omega^{2}(q)}
  \left(C_B^{q}\, A_C^{q}- A_B^{q}\,C_C^{q} \right)^{2}.
\end{equation}

Fig.~5 shows $\mathcal N_(q,\tau_M)$ as a function of the rescaled quasi-momentum $q/N$, treated as a continuous variable, for different values of $n$. Panel (a) displays a sequence of curves corresponding to a favorable scenario exhibiting finite-size super-extensive scaling, while panel (b) illustrates an unfavorable case. A clear trend emerges: As $n$ increases, the occupation numbers $\mathcal N_(q,\tau)$ decrease for most quasi-momenta and develop a larger number of oscillations. This behavior originates from the combined effect of multiple harmonics and long-range hopping processes.

Finite-size super-extensive scaling is therefore achieved when, upon increasing the number of spins $N$—that is, as the quasi-momentum sampling becomes denser—the parameter $n$ decreases.
\begin{figure}[H]
    \centering
    \begin{subfigure}{0.5\textwidth}
        \centering
        \includegraphics[width=\linewidth]{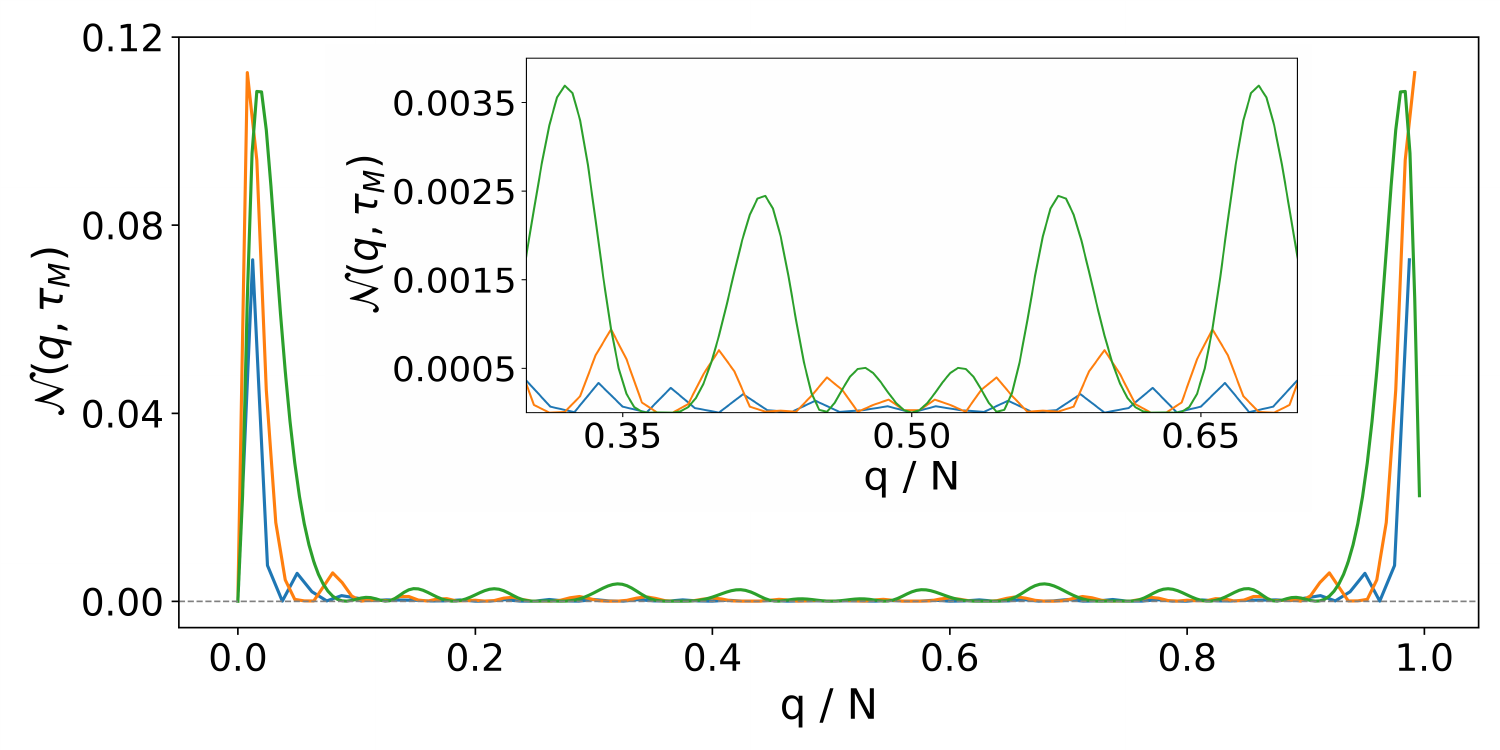}
        \caption{}
    \end{subfigure}
    \hfill
    \begin{subfigure}{0.5\textwidth}
        \centering
        \includegraphics[width=\linewidth]{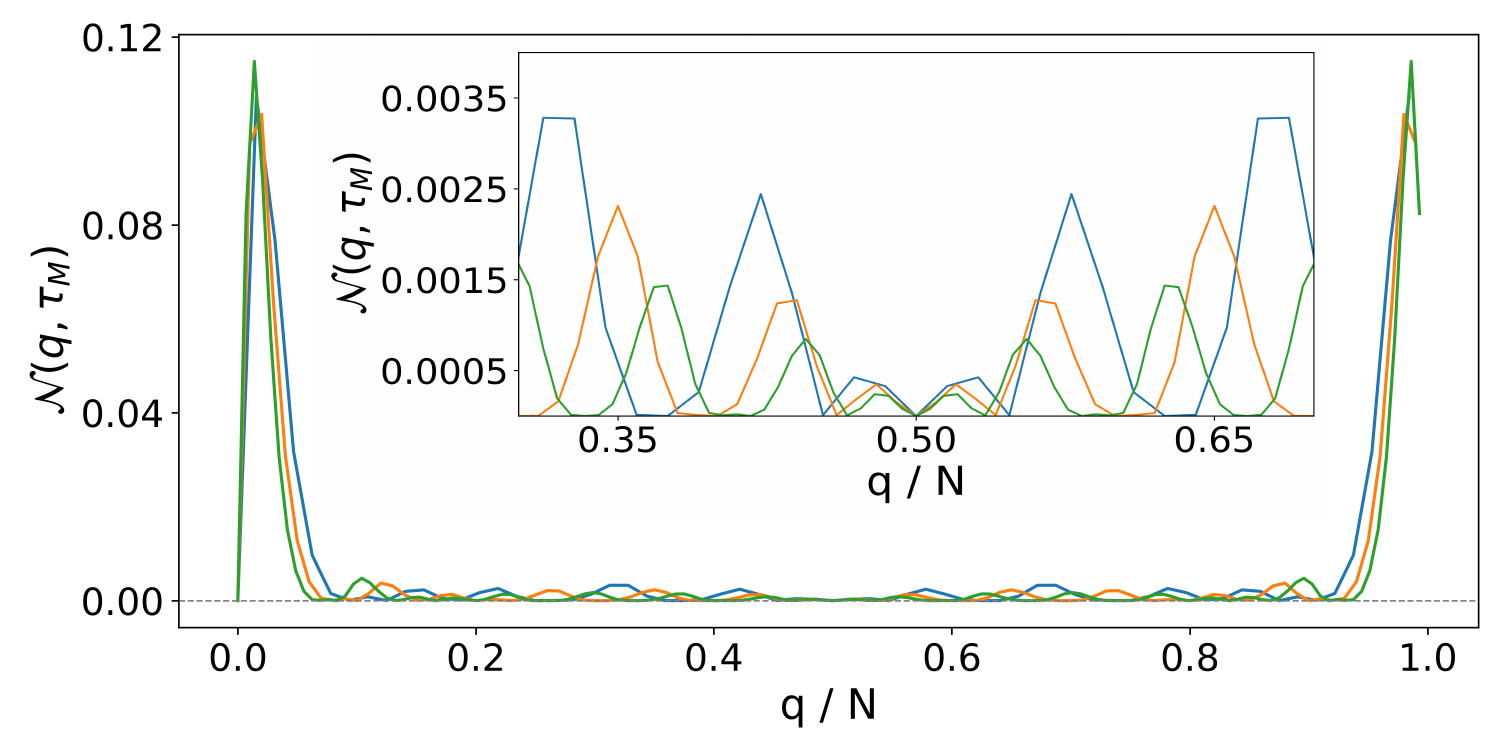}
        \caption{}
    \end{subfigure}
    
    \caption{(a) Fermionic occupation number $\mathcal{N}_q$ as a function of $q/N$ for $\phi_B = 0.6$ and $\phi_C = \pi/4$, with fixed product $nN = 2000$. Results are shown for $n = 25$ (blue), $n = 16$ (orange), and $n = 8$ (green). \\
(b) Same as in (a), but for $n = \sqrt{N}$, with $n = 8$ (blue), $n = 10$ (orange), and $n = 12$ (green). Insets show a zoom of the central region.}
    \label{Num_Occ_Plots}
\end{figure}

{\color{black}
\section{Unfavorable model} \label{sec:unfavorable}
Now, we investigate the favorable regimes discussed in the previous section in another kind of generalized cluster-Ising model. This new Hamiltonian reads
\begin{equation}
    \widetilde{H}=- \cos\widetilde{\phi} \sum_{j=1}^{N} 
\sigma^{x}_{j}\, O^{z}_{j,n}\, \sigma^{x}_{j+n+1}
+ \sin\widetilde{\phi} \sum_{j=1}^N 
\sigma^{z}_{j} \label{eq:H1}
\end{equation}
where $O^{z}_{j,n}$ is defined in Eq.\eqref{Ozjn}, while $\widetilde{\phi}$ is a free real parameter. The model is exactly solvable like the one analyzed in the main text, but differently from the main text's Hamiltonian, here a single term with a string of $n+1$ Pauli matrices is present. Following the same steps used to diagonalize $H$, we can obtain for this Hamiltonian as well the form of Eq. \eqref{Ham_With_Tau_Matrices} after defining
\begin{equation}
    \begin{aligned}
        & \widetilde{A}^q = \cos\!\left( \frac{2\pi}{N} (n+1) q \right)\cos\widetilde{\phi}
 - \sin\widetilde{\phi} \\
&\widetilde{C}^q = - \sin\!\left( \frac{2\pi}{N} (n+1) q \right)\cos\widetilde{\phi}.
    \end{aligned}
\end{equation}
As shown in Fig. \ref{H_Tilde_Power}, $nN = $ constant, which is a favorable regime for $H$, does not show a super-extensive scaling in the maximum charging power $\widetilde{P}_M$ for the system described by $\widetilde{H}$. The same conclusion can be stated for the other regimes (not shown).

\begin{figure}[H]
    \centering
    \includegraphics[width=0.9\linewidth]{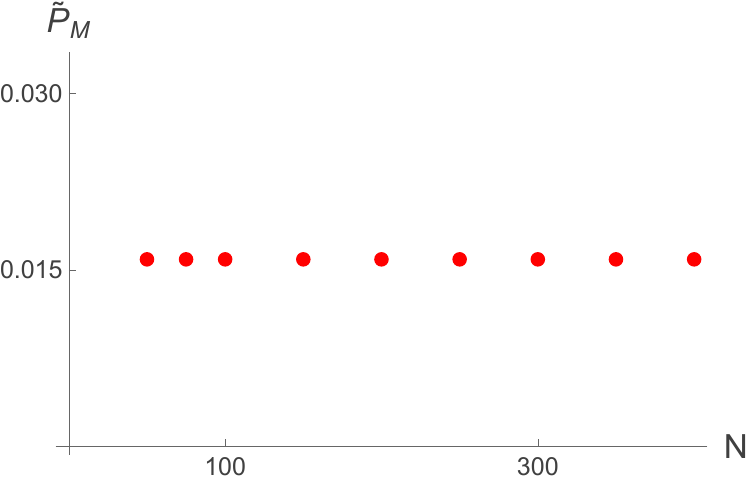}
    \caption{Plot of the maximum charging power $\widetilde{P}_M$ for the system $\widetilde{H}$ as function of $N$ for $\phi_B = 0.6$, $\phi_C = \pi/4$ and zero temperature. The reported case is $nN = $ 2000.}
    \label{H_Tilde_Power}
\end{figure}
If we now perform the same analytical study we reported in Sec. \ref{subsec_short_time} by considering a sudden double quench from $\widetilde{\phi}_B = 0$ to $\widetilde{\phi}_c = \widetilde{\phi}$, we have
\begin{equation}
    \left(\widetilde{C}_B^{q}\, \widetilde{A}_{C}^{q} - \widetilde{A}_{B}^{q}\,\widetilde{C}_{C}^{q} \right)^{2} = \sin^2\left(\frac{2\pi(n+1)}{N}q\right) \sin^2(\widetilde{\phi}),
\end{equation}
while $\varepsilon_1(q) = 1$. So, the slope $\gamma$ becomes
\begin{equation}
    \gamma = \frac{\sin^2(\widetilde{\phi})}{N}\sum_{q=0}^{N-1} \sin^2\left(\frac{2\pi(n+1)}{N}q\right) =\frac{\sin^2(\widetilde{\phi})}{2}
\end{equation}
where we used the identity
\begin{equation*}
    \sum_{q=0}^{N-1} \sin^2\left(\frac{2\pi(n+1)}{N}q\right) = \frac{N}{2}.
\end{equation*}
So, in the short-time limit, the slope of the charging power for this new model depends only on the $\widetilde{\phi}$ parameter and not on the number of sites $N$. This is a relevant observation to explain why the super-extensive trend does not emerge in this model.\\ In the present case the number of the oscillations of $\mathcal N_q$ increases with the increase of $n$, but their amplitude remains qualitatively the same.

\section{Conclusions} \label{conclusions}
We have investigated the charging performance of quantum batteries based on a \textcolor{black}{Jordan-Wigner} integrable extensions of the cluster–Ising spin chain. Despite the integrability of the model, we have shown that it can exhibit an apparent super-extensive scaling of the charging power across wide intervals of system sizes, reaching up to $O(10^3)$ spins. Our analysis demonstrates that this behavior does not originate from an anomalous dependence of the optimal charging time on the system size, but rather from a super-extensive growth of the stored energy itself. As such, the effect cannot persist in the thermodynamic limit and should instead be interpreted as a finite-size enhancement. We verified that the phenomenon is robust against thermal effects, which affect quantitative values but not the qualitative trends. Subsequently, we have deepened the analysis by assessing in a semi-analytical way the behavior at short times, showing that the super-extensive behavior is already visible in the initial slope of the charging power. Finally, to try to isolate the necessary ingredients for the phenomenon and hence to address possible extensions, we have analyzed the fermionic occupation numbers. What emerges is that to observe the finite-size super-extensive scaling the crucial ingredient is to decrease the number and increase the amplitude of the oscillations in the occupation numbers as a function of quasi-momentum as the 'sampling', that is the number of spins $N$, increases. We argue that this property can be achieved when both long range hopping and multiple harmonics characterize the model. In the last Section we show that when a single long range hopping term is present the effect is absent by briefly analysing an unfavorable model where the finite-size super-extensive charging power is not observed.

Altogether, our results highlight that even fermionic integrable models—typically considered incapable of supporting enhanced charging power under quench protocols~\cite{JuliaFarrePRR2020}—can display strong finite-size effects mimicking super-extensive performance. These findings can help delineate the boundaries between genuine collective advantages and finite-size artifacts in quantum battery architectures.

\appendix

\section{Derivation of Eq. \eqref{Energy_Chaos}} \label{App:Energy_Chaos}
In this Appendix, we provide more details about the calculation that leads to Eq. \eqref{Energy_Chaos}, presenting the main steps that allow to derive that formula. The first step consists in diagonalizing the Hamiltonian in Eq. \eqref{Ham_Sum_To_Diag}. In order to do so, we have to introduce a new set of fermions such that
\begin{equation}\label{Canonic_Transf}
    \boldsymbol{\xi}_{B,C} = U_{B,C} \boldsymbol{\psi}
\end{equation}
where  $\boldsymbol{\xi} = (\xi_q,\xi^\dag_{-q})^T$, $\boldsymbol{\psi} =(\psi_q,\psi^\dag_{-q})^T$, $U_{B,C}$ is a unitary $2\times 2$ matrix, while subscript $B$ ($C$) indicates that we are referring to the QB (charging) Hamiltonian. Let's consider
\begin{equation} \label{UBC}
    U_{B,C} = \frac{1}{\sqrt{2\varepsilon_{B,C}}} \begin{pmatrix}
        - \frac{C_{B,C}^q}{\sqrt{\varepsilon_{B,C} + A_{B,C}^q}} & \sqrt{\varepsilon_{B,C} + A_{B,C}^q}\\
        \frac{C_{B,C}^q}{\sqrt{\varepsilon_{B,C} - A_{B,C}^q}} &\sqrt{\varepsilon_{B,C} - A_{B,C}^q}
    \end{pmatrix}.
\end{equation}
With reference to the main text, here we set $\varepsilon_{B}=\epsilon$, $\varepsilon_{C}=\omega$. It is possible to show that $U_{B,C}$ satisfies three properties: it is unitary, it makes the transformation reported in Eq. \eqref{Canonic_Transf} canonical, and it satisfies
\begin{equation*}
    U_{B,C} \mathcal{H}_{B,C}^{q}U_{B,C}^\dagger = \begin{pmatrix}
        \varepsilon_{B,C} &0 \\
        0 &-\varepsilon_{B,C}
    \end{pmatrix}.
\end{equation*}
Via Eq. \eqref{Canonic_Transf}, one can establish a relation between the fermionic operators that diagonalize the QB Hamiltonian and those that diagonalize the charging Hamiltonian. This relation reads
\begin{equation} \label{xi_B_M_xi_C}
    \boldsymbol{\xi}_{B} = U_{B} \boldsymbol{\psi} = U_{B} (U_C^{-1} \boldsymbol{\xi}_{C}) = M\boldsymbol{\xi}_{C},
\end{equation}
where $M \equiv U_{B} U_C^{-1}$. In terms of the $\boldsymbol{\xi}_{B}$ fermions, the QB Hamiltonian takes the diagonal form
\begin{equation} \label{Initial_Hamiltonian_xi_B}
    H_B = \sum_q \varepsilon(q) \left(\xi_{B,q}^\dagger \xi_{B,q} + \xi_{B,-q}^\dagger \xi_{B,-q}\right).
\end{equation}
From here, it's possible to compute the initial energy of the QB by averaging over the initial thermal state $\rho$. This reads
\begin{equation} \label{App_Initial_Energy}
\begin{aligned}
    \langle H_B\rangle &= \sum_q \varepsilon(q)\left(\langle\xi^\dagger_{B,q}\xi_{B,q}\rangle
    +
    \langle \xi^\dagger_{B,-q}\xi_{B,-q}\rangle
    \right) = \\&= 2 \sum_q \varepsilon(q)n_F(q),
\end{aligned}
\end{equation}
with $n_F(q) = (e^{\beta \varepsilon(q)}+1)^{-1}$. We now consider the time evolution generated by the charging Hamiltonian $H_C$, which is diagonal in the $\boldsymbol{\xi}_C$ basis. By explicitly writing Eq.~\eqref{xi_B_M_xi_C}, one obtains
\begin{equation}
    \begin{cases}
        \xi_{B,q} = M_{11} \xi_{C,q} + M_{12} \xi^\dagger_{C,-q}, \\ 
        \xi^\dagger_{B,-q} = M_{21} \xi_{C,q} + M_{22} \xi^\dagger_{C,-q}.
    \end{cases}
\end{equation}
Using these relations, $H_B$ can be expressed in terms of the $\boldsymbol{\xi}_C$ fermions, allowing us to evaluate its time evolution due to the sudden quench. This yields
\begin{equation} \label{HB_With_Xi_C}
\begin{aligned}
    H_B(t)=&\left(|M_{11}|^2 - |M_{21}|^2\right) \xi^\dagger_{C,q}\xi_{C,q} \\&+ \left(|M_{22}|^2 - |M_{12}|^2\right) \xi^\dagger_{C,-q}\xi_{C,-q} \\& + \left(M_{11}^*M_{12} - M_{21}^* M_{22}\right) \xi^\dagger_{C,q}\xi^\dagger_{C,-q} e^{2i\omega(q) t} \\&+ \left(M_{12}^*M_{11} - M_{22}^* M_{21}\right) \xi_{C,-q}\xi_{C,q} e^{-2i\omega(q) t}.
\end{aligned}
\end{equation}
Finally, we need to rewrite Eq. \eqref{HB_With_Xi_C} in terms of the original fermionic operators by exploiting the relation $ \boldsymbol{\xi}_{C} = M^\dagger \boldsymbol{\xi}_{B}$. At this stage we observe that, since our final goal is to compute the expectation value of the energy over a thermal state, all terms involving fermionic operators with different quasi-momenta can be safely neglected. More explicitly, for $q \neq q'$ one has
\begin{equation}
    \langle \xi^\dagger_{q} \xi_{q'} \rangle = 0, \qquad 
    \langle \xi_{q} \xi_{q'} \rangle = 0, \qquad
    \langle \xi^\dagger_{q} \xi^\dagger_{q'} \rangle = 0.
\end{equation}
As a consequence, only terms that are diagonal in momentum space give a non-vanishing contribution to the energy expectation value. Once $H_B(t)$ has been expressed in terms of the $\xi_B$ fermions, the stored energy is obtained by taking the expectation value of $H_B(t)$ over the thermal state $\rho$, and subtracting the initial energy given in Eq. \eqref{App_Initial_Energy}. This procedure ultimately leads to Eq. \eqref{Energy_Chaos} of the main text. For example, the oscillatory cosine term originates from the complex exponentials $\exp\{\pm 2i\omega(q)t\}$ appearing in Eq. \eqref{HB_With_Xi_C}, while the temperature dependence arises from expressions of the form
\begin{equation}
    1-2n_F(q) = 1-\frac{2}{e^{\beta \varepsilon(q)}+1} = \frac{e^{\beta \varepsilon(q)}-1}{e^{\beta \varepsilon(q)}+1} = \tanh\left(\frac{\beta \varepsilon(q)}{2}\right),
\end{equation}
which naturally emerge from combinations of fermionic expectation values such as $\langle \xi^\dagger_{B,q} \xi_{B,q} \rangle$ and $\langle \xi_{B,-q} \xi^\dagger_{B,-q} \rangle$, and are further simplified by exploiting the unitarity of the $M$ matrix, which enforces nontrivial cancellations among coefficients.

\section{Density plots of other favorable cases}
Fig. \ref{App_DP} shows the density plots of the fit coefficients $a$ and $\alpha$ for the other favorable cases discussed in the main text, corresponding to the scalings $n\sqrt{N}=\mathrm{constant}$ and $n\sqrt[3]{N}=\mathrm{constant}$.
\begin{figure}[H]
    \centering
    
    \begin{subfigure}{0.45\columnwidth}
        \centering
        \includegraphics[width=\linewidth]{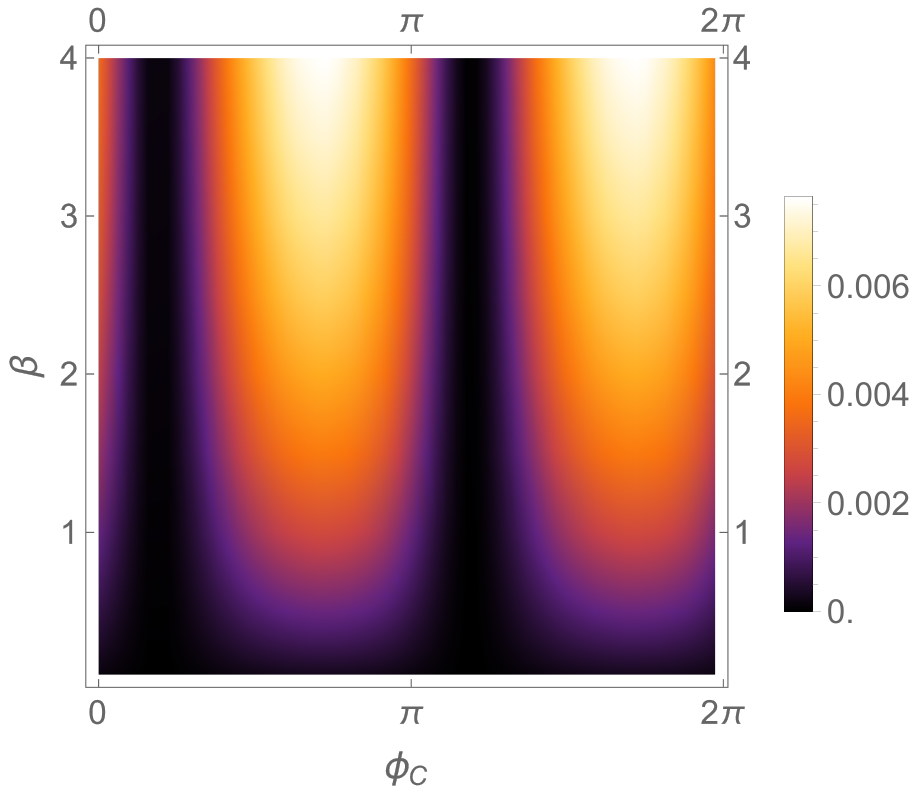}
        \caption{}
    \end{subfigure}
    \hfill
    \begin{subfigure}{0.45\columnwidth}
        \centering
        \includegraphics[width=\linewidth]{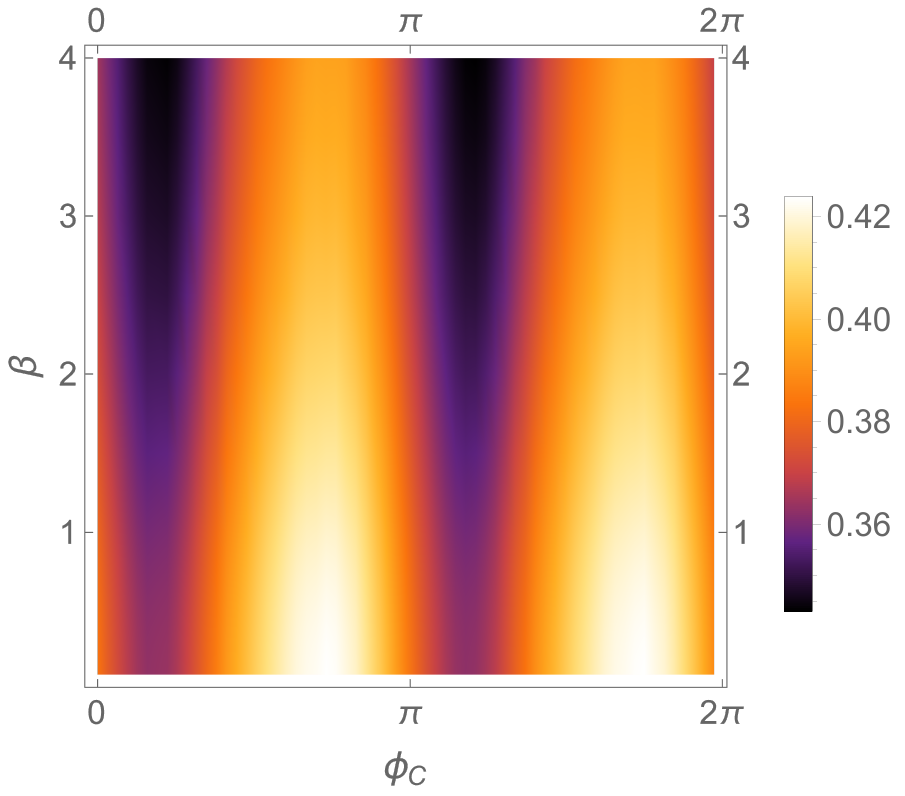}
        \caption{}
    \end{subfigure}

    \vspace{0.5cm}

    \begin{subfigure}{0.45\columnwidth}
        \centering
        \includegraphics[width=\linewidth]{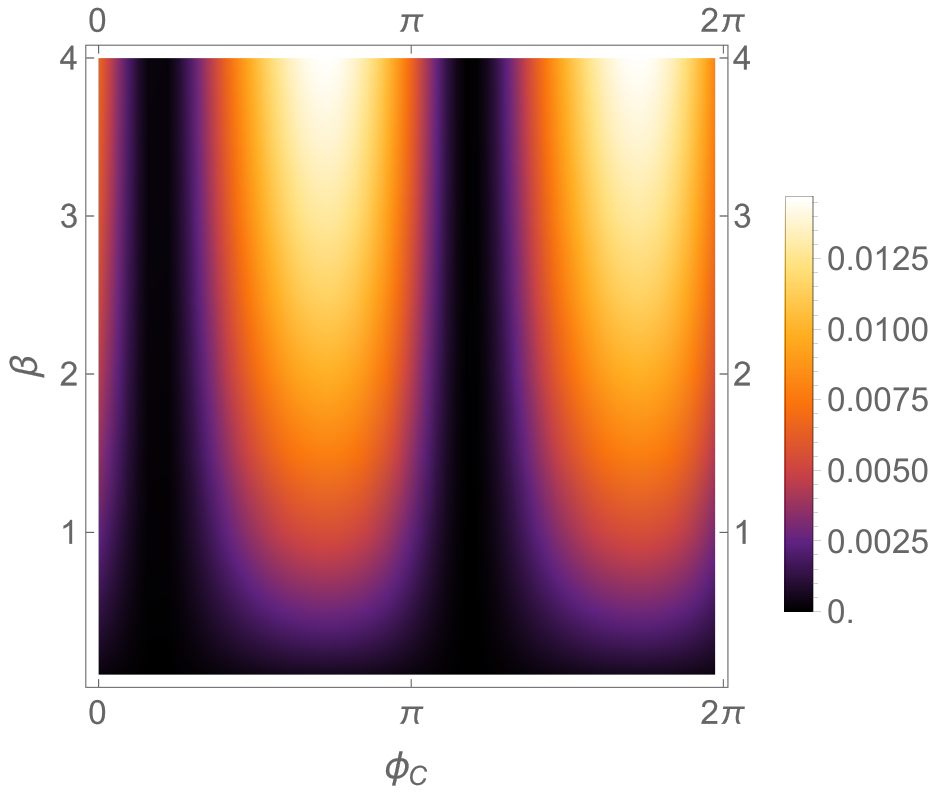}
        \caption{}
    \end{subfigure}
    \hfill
    \begin{subfigure}{0.45\columnwidth}
        \centering
        \includegraphics[width=\linewidth]{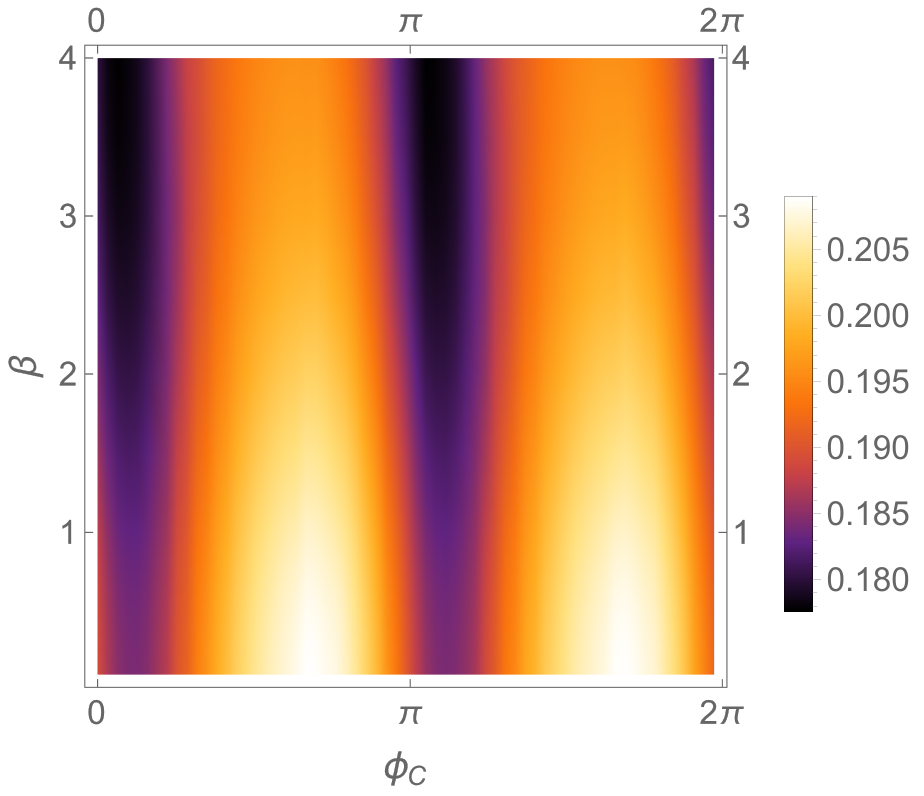}
        \caption{}
    \end{subfigure}

    \caption{Density plots of the fit coefficients for the scalings $n\sqrt{N}= 100$ (coefficient $a$ in panel (a), coefficient $\alpha$ in panel (b)) and $n\sqrt[3]{N}= 50$ (coefficient $a$ in panel (c), coefficient $\alpha$ in panel (d)), with fixed $\phi_B=0.6$.}
    \label{App_DP}
\end{figure}

}
}
\begin{acknowledgements} D.F. acknowledge support from the project PRIN 2022 - 2022XK5CPX (PE3) SoS-QuBa - "Solid State Quantum Batteries: Characterization and Optimization" funded within the programme "PNRR Missione 4 - Componente 2 - Investimento 1.1 Fondo per il Programma Nazionale di Ricerca e Progetti di Rilevante Interesse Nazionale (PRIN)", funded by the European Union - Next Generation EU".
\end{acknowledgements}

\end{document}